\documentclass[12pt,noshowpacs,nofootinbib,notitlepage,amsmath,amssymb]{revtex4-2}
\usepackage{setspace}
\usepackage[top=1in,bottom=1in,left=1in,right=1in]{geometry}
\usepackage{graphicx,color}
\usepackage[colorlinks=true,citecolor=blue,linkcolor=blue,urlcolor=blue]{hyperref}
\usepackage{bm}
\hypersetup{
  pdftitle={Superselected ghost theory: entangled pairs},
  pdfauthor={Bob Holdom}
}

\begin{document}

\title{\color{blue}\Large Superselected ghost theory: entangled pairs}

\author{Bob Holdom}
\email{bob.holdom@utoronto.ca}
\affiliation{Department of Physics, University of Toronto\\ Toronto, Ontario, Canada M5S 1A7}

\begin{abstract}
The superselection-rule approach to ghost theories is extended to theories
with a nonreal spectrum containing a complex-conjugate pair of poles on the
physical sheet. The two excitations are labeled by their respective complex
masses, $M$ and $M^*$, and the superselection sectors are labeled by $n$, the
difference between the numbers of $M$ and $M^*$ excitations. The corresponding
generalized ghost parity $Q$ assigns a phase $e^{in\alpha}$ to each sector. Under $Q$
superselection, nonvanishing norms occur only in the $n=0$ sector, and
the physical state condition $Q^{2}|s\rangle=|s\rangle$ projects
onto this sector. For a $MM^*$ pair carrying real total energy and momentum,
we define a swap operation $R$. The $R$-eigenstates are entangled pairs, and an $R$
superselection rule ensures positive probabilities. We discuss how the optical
theorem describes physical cuts through these entangled pairs.
\end{abstract}

\maketitle

\section{Introduction}\label{s1}

This paper is a companion to \cite{realsp}, where ghost theories with a
real spectrum were analyzed through a superselection rule associated with an
exact ghost-parity symmetry $Q$. In that work, we described the emergence of a
probability interpretation and a consistent optical theorem. The spectral
representation of the propagator decomposed sector by sector and retained
the standard analytic structure. Consequently, complex-conjugate poles do not
appear on the physical sheet. A transformed perturbation theory that modifies
old-fashioned perturbation theory realizes these results \cite{pertur}. The
real-spectrum regime thus appears to be self-consistent in
the sense that perturbative corrections no longer imply complex-conjugate
poles. However, it is still of interest to consider the complementary regime
in which the exact spectrum
contains a pair of complex-conjugate energies, perhaps due to the
merging of two single-particle real-energy levels, one ghost and one non-ghost,
for sufficiently large coupling.

Complex-conjugate poles on the physical sheet have motivated various
prescriptions intended to reconcile them with unitarity and causality. The
Lee-Wick construction \cite{LW:1969}  defines a projected,
future-boundary $S$-matrix prescription that removes exponentially growing modes.
In \cite{Grinstein:2007mp,Donoghue:2019ecz}, a non-Feynman contour implements a backward-in-time
prescription that avoids the exponential growth.
The fakeon prescription \cite{Anselmi:2018kgz} removes the ghost from the
asymptotic spectrum entirely through a different choice of Green's function.
Meanwhile, \cite{Kubo:2023lpz,Kubo:2024ysu} emphasize
that these various prescriptions are inconsistent with standard canonical quantization.
A ghost-masking effect within the context of canonical quantization has been studied
in~\cite{Buoninfante:2026mve,Buoninfante:2026jfj}.

Here we explore whether superselection can provide a way to reconcile canonical
quantization with a probability interpretation. Our construction
has two stages. First, the single-particle states associated
with the masses $M$ and $M^*$ have vanishing norms but a nonzero off-diagonal
inner product. The operator $Q$, the analog of ghost
parity, assigns conjugate phases, rather than opposite signs, to these two
states. The resulting superselection sectors are labeled by the integer
$n=N_M-N_{M^*}$. Under $Q$ superselection, superpositions between sectors are excluded, and
only states in the $n=0$ sector have nonvanishing norm. A physical state condition
$Q^{2}|s\rangle=|s\rangle$ eliminates all other sectors.

In the center-of-momentum (CoM) frame, an $MM^*$ pair whose constituents have real
spatial momenta $\mathbf k$ and $-\mathbf k$ and complex-conjugate energies has
real total energy.
In any frame, the pair has a real 4-momentum with an invariant mass that varies continuously with
$|\mathbf k|$. Within this pair space, a $\mathbb Z_2$ operation $R$ swaps the
assignments of $\mathbf k$ and $-\mathbf k$ to $M$ and $M^*$. The $R$-even and
$R$-odd combinations are entangled states with positive and negative norms, respectively.
We use $R$ to impose another superselection rule on the theory. As with $Q$,
this organizes the positive- and negative-norm states into two distinct
sectors, thereby ensuring that all probabilities are positive.

Unitarity is illustrated in old-fashioned perturbation theory, where a pair contributes
to the real energy of an intermediate state. The discontinuity of the resulting energy denominator
defines a phase-space measure that describes a continuum of real invariant
masses for the pair's total momentum. The optical theorem is recovered by using
the pair completeness relation and the residue of the pair propagator. The
physical cut has a positive weight and passes through an entangled
$MM^*$ pair treated as a single composite intermediate excitation.

Section~\ref{s2} identifies $Q$, the ghost phase operator, and
Section~\ref{s3} applies it as a superselection rule, leading to a pair
propagator and pair kinematics. Section~\ref{s4} introduces the swap symmetry
$R$, identifies entangled pair states as the $R$-eigenstates, and obtains the
pair-state completeness relation. Section~\ref{s5} sketches an old-fashioned
perturbation theory formulated directly in terms of the physical pair states
and then discusses the pair contribution to the optical
theorem. Section~\ref{s6} fixes the
normalization, derives the phase-space measure, and presents a self-energy
example. Section~\ref{s7} concludes.

\section{Ghost phase operator}\label{s2}

We use the matrix notation of \cite{realsp} to extend the analysis beyond the
real-spectrum regime. A pair of
complex-conjugate poles on the physical sheet is characterized by $p^2=M^2$ and
$p^2=M^{*2}$, where $M$ is complex. We represent the energy eigenstates by
the column vectors $\bm\varphi^{M}$ and $\bm\varphi^{M*}$, with the spatial momentum
$\mathbf{p}$ left implicit. With the matrix $\bm\eta$ defining the native inner
product, the single-particle
norms $\bm{\varphi}^{M\dagger}\bm\eta\bm{\varphi}^M$
and $\bm{\varphi}^{M^*\dagger}\bm\eta\bm{\varphi}^{M^*}$ vanish. Instead, we have the
``off-diagonal norm''
$\bm{\varphi}^{M^*\dagger}\bm\eta\bm{\varphi}^M
=(\bm{\varphi}^{M\dagger}\bm\eta\bm{\varphi}^{M^*})^*$. This more general structure
is compatible with unitarity in the presence of an indefinite inner product.

The dual projectors
\begin{align}
\mathbf P_M(\mathbf p)&=
  \frac{\bm{\varphi}^M\bm{\varphi}^{M^*\dagger}\bm\eta}
       {\bm{\varphi}^{M^*\dagger}\bm\eta\bm{\varphi}^M},&
\mathbf P_{M^*}(\mathbf p)&=
  \frac{\bm{\varphi}^{M^*}\bm{\varphi}^{M\dagger}\bm\eta}
       {\bm{\varphi}^{M\dagger}\bm\eta\bm{\varphi}^{M^*}},
\end{align}
satisfy
\begin{align}
\mathbf P_M\bm\varphi^M=\bm\varphi^M,\qquad
\mathbf P_{M^*}\bm\varphi^{M^*}=\bm\varphi^{M^*},\qquad
\mathbf P_M\bm\varphi^{M^*}=\mathbf P_{M^*}\bm\varphi^M=0.
\end{align}
$\tilde{\mathbf{H}}=\bm\eta\mathbf H$ is the matrix representation of the
pseudo-Hermitian Hamiltonian that generates time evolution, and $\mathbf{Q}$ is the matrix
representation of the superselection charge. The single-particle contributions
from $M$ and $M^*$ to the completeness relation and to these two operators are
then
\begin{align}
\bm1&\ni\int d^3\mathbf{p} \left(
\mathbf P_M+\mathbf P_{M^*}
\right),\label{e2}\\
\tilde{\mathbf{H}}&\ni\int d^3\mathbf{p} \left(
E_{\mathbf p}\mathbf P_M+E_{\mathbf p}^*\mathbf P_{M^*}
\right),\\
\mathbf{Q}&\ni\int d^3\mathbf{p} \left(
e^{i\alpha}\mathbf P_M+e^{-i\alpha}\mathbf P_{M^*}
\right),
\label{e3}\end{align}
where $E_{\mathbf p}=\sqrt{\mathbf p^2 + M^2}$. In particular,
\begin{align}
\mathbf{Q}\bm\varphi^M=e^{i\alpha}\bm\varphi^M,\qquad
\mathbf{Q}\bm\varphi^{M^*}=e^{-i\alpha}\bm\varphi^{M^*}.
\end{align}
We take $e^{i\alpha}$ to be the phase of
$\bm{\varphi}^{M^*\dagger}\bm\eta\bm{\varphi}^M$.
In the real-spectrum case, the norms are positive or negative, and their absolute
values were used in the definition of $\mathbf{Q}$ so that its action on a
state gives the sign of the state's norm. Similarly, the phases
in (\ref{e3}) are absent when the projector denominators are replaced by
$|\bm{\varphi}^{M^*\dagger}\bm\eta\bm{\varphi}^M|$. $e^{i\alpha}$ is dynamically determined
once a phase convention is established, and its only property of relevance
is that $\alpha/(2\pi)$ be irrational.

If there were more than one complex-conjugate pair, the two operators would
receive additional contributions with different complex masses and phases.
The following discussion can be generalized accordingly.

For an operator $\mathbf A$, self-adjointness with respect to the $\eta$ inner
product means $\mathbf A^\dagger\bm\eta=\bm\eta\mathbf A$. Both $\tilde{\mathbf H}$ and $\mathbf Q$ are self-adjoint. Because they are
defined using the same set of projectors, they also satisfy
$[\mathbf Q,\tilde{\mathbf H}]=0$. The integer $n = N_M - N_{M^*}$ serves as
the fundamental conserved superselection charge; it is the difference between
the occupation numbers of $M$ and $M^*$ in a state.
$\mathbf Q$ acts on such a state by multiplication by the ghost phase
$e^{in\alpha}$. In contrast to the
real-spectrum case, the property $\mathbf{Q}^2=1$ no longer holds.

For real $\beta$, we may consider $e^{i\beta\mathbf{Q}}$, which
also commutes with $\tilde{\mathbf{H}}$. Because $\mathbf{Q}$ is
self-adjoint, transforming the states by
$e^{i\beta\mathbf{Q}}$ leaves the inner product invariant. On a sector with
ghost phase $e^{in\alpha}$, this transformation acts as
$e^{i\beta e^{in\alpha}}=e^{i\beta\cos(n\alpha)}e^{-\beta\sin(n\alpha)}$,
thereby inducing exponential amplification or suppression of states in the
$n\neq0$ sectors.
Time evolution generated by $e^{-i\tilde{\mathbf{H}}t}$ similarly produces the
exponential factor $e^{{\rm Im}(E_{\rm tot})t}$, where ${\rm Im}(E_{\rm tot})$ is
generally nonzero and of either sign for the $n\neq0$ sectors.

We now switch from matrix and column-vector notation and denote
the energy eigenstates by $|M_p\rangle$ and
$|M_{p^{*}}^*\rangle$. They are labeled by their on-shell 4-momenta, which
are determined by their 3-momenta. The off-diagonal norm is
$\langle M^*_{p^{*}} | M_p\rangle
=\bm{\varphi}^{M^*\dagger}\bm\eta\bm{\varphi}^M
\equiv\langle M^* | M\rangle$. We assume a convention where
$\langle M^* | M\rangle$ does not depend on $|\mathbf{p}|$. A multiparticle state
can have a nonzero inner product with itself only if each $M$ excitation can be
paired with an $M^*$ excitation, and vice versa. Any state characterized by a
$n\neq0$ therefore has zero norm.
Superpositions of states from different sectors can have nonzero norms. For
example, $|M_p\rangle\pm|M^{*}_{p^{*}}\rangle$ has the norm
$\pm2\mathrm{Re}[\langle M^*|M\rangle]$.

\section{$Q$ superselection rule}\label{s3}

When $Q$ is imposed as a superselection rule, superpositions of states from
different sectors are forbidden, and nonvanishing norms occur only in the $n=0$
sector. In addition, we impose the physical-state condition
$Q^{2}|s\rangle=|s\rangle$. For irrational $\alpha/(2\pi)$, this condition
projects out all $n\neq0$ sectors. The $Q$ superselection structure for normal
ghosts, which have real masses, remains as described in \cite{realsp}.
The lightest normal ghost is stable.
States with $Q=1$ contain an even number of normal ghosts and
any number of $MM^{*}$ pairs; states with $Q=-1$ contain an odd number of
normal ghosts and any number of $MM^{*}$ pairs.

Let $\phi$ be a fundamental Hermitian field that can create an $M$ or $M^*$
excitation from the vacuum. The corresponding propagator $\Delta(p)$ receives
the following contributions near the complex-conjugate
poles~\cite{Holdom:2024onr},
\begin{align}
\Delta(p)\big|_{p^2 \to M^2} &\to
\frac{\langle  0|\phi|M\rangle
\langle M^*|\phi| 0\rangle}
{\langle M^*|M\rangle}
\frac{i}{p^2 - M^2 + i\epsilon},\nonumber \\
\Delta(q)\big|_{q^2 \to M^{*2}} &\to
\frac{\langle  0|\phi|M^*\rangle
\langle M|\phi| 0\rangle}
{\langle M|M^*\rangle}
\frac{i}{q^2 - M^{*2} + i\epsilon}.
\label{e5}\end{align}
The residues reflect the off-diagonal nature of the projectors
$\mathbf P_M$ and $\mathbf P_{M^*}$ in (\ref{e2}). If $Z$ and $Z^{*}$ denote the two residues, then
$Z+Z^{*}$ is negative if $\phi$ is a ghost field.

This two-pole structure can be mimicked by the free effective Lagrangian
\cite{naka,Kubo:2024ysu}
\begin{align}
\mathcal{L}_{\rm free} = \frac{1}{2} \left[ \partial_\mu \varphi \partial^\mu \varphi - M^2 \varphi^2 + \partial_\mu \varphi^\dagger \partial^\mu \varphi^\dagger - M^{*2} (\varphi^\dagger)^2 \right].
\label{e1}\end{align}
When defined using the $p^{0}$ integration contours implied by canonical
quantization \cite{Kubo:2024ysu}, the resulting propagators
$Z\langle0|T[\varphi(x)\varphi(y)]|0\rangle$ and
$Z^{*}\langle0|T[\varphi^{\dagger}(x)\varphi^{\dagger}(y)]|0\rangle$ are the
coordinate-space versions of those in (\ref{e5}).
Although the Lagrangian $\mathcal{L}_{\rm free}$ contains terms that change $n$
by two units, the $Q^{2}|s\rangle=|s\rangle$ condition acts to project the
$S$-matrix onto the physical $n=0$ space.
This projection removes unpaired $M$ and $M^{*}$ excitations and leaves a $MM^{*}$ pair
propagator equal to the product of the propagators in (\ref{e5}).
We use its residue in Section~\ref{s5}.\footnote{
$\mathcal{L}_{\rm free}$ also defines a minimal coupling to gravity and thus
determines the minimal couplings between gravitons and $MM^{*}$ pairs.
With such interactions, a transformed perturbation theory, as in~\cite{realsp,pertur}, is
required so that the superselection structure is preserved order by order.}

We further require the total 4-momentum $P$ of each $MM^*$ pair to be real.
This requirement is most easily understood in the
CoM frame, where the spatial momenta of $M$ and $M^*$ are
$\mathbf{k}$ and $-\mathbf{k}$, respectively. With real $\mathbf{k}$
and $E_{\mathbf k}=\sqrt{\mathbf k^2 + M^2}$, the momenta are
\begin{align}
&p_{\rm CoM}=(E_{\mathbf k},\mathbf k),\qquad
q_{\rm CoM}=(E_{\mathbf k}^*,-\mathbf k),\\
&P_{\rm CoM}=p_{\rm CoM}+q_{\rm CoM}=(\mu(|\mathbf k|),\mathbf0),
\end{align}
where the real invariant mass of the pair is
\begin{align}
\mu(|\mathbf k|)=E_{\mathbf k}+E_{\mathbf k}^{*}.
\label{epairmass}
\end{align}
Under a real boost to any other frame,
the generally complex 4-momenta $p$ and $q$ satisfy
\begin{align}
p^2=M^2,\qquad q^2=M^{*2},\qquad p+q=P.
\end{align}
In any frame, the real total 4-momentum is
$P^{\mu}=(\Omega(\mathbf P,\mathbf k),\mathbf P)$, where
\begin{align}
\Omega(\mathbf P,\mathbf k)
=\sqrt{\mathbf P^2+\mu^2(|\mathbf k|)}.
\label{epairenergy}
\end{align}
With $p$ and $q$ related to $\mathbf k$ and $\mathbf P$ in this way, we can write a pair state as
$|M_p M^*_q\rangle$. Because $q$ and $p$ have equal and opposite imaginary
parts, the exponential time dependences of the two coordinate-space propagators
mentioned above cancel in their product.

\section{$R$ and another superselection rule}\label{s4}

We introduce a $\mathbb{Z}_2$ swap operation
\begin{align}
R|M_p M^*_q\rangle
=|M_{q^*} M^*_{p^*}\rangle.
\end{align}
Because $R^2 = 1$, $R$ is effectively an internal parity-like operator for the
composite two-particle state.
Since a Lorentz transformation $\Lambda$ is real, $(\Lambda q)^*=\Lambda q^*$.
It follows that the swap on the momentum labels
commutes with Lorentz transformations,
\begin{align}
R(\Lambda p,\Lambda q)
=((\Lambda q)^*,(\Lambda p)^*)
=\Lambda(q^*,p^*)
=\Lambda R(p,q).
\end{align}
The transformed momenta remain on the appropriate mass shells and preserve
the total momentum,
\begin{align}
(q^*)^2=M^2,\qquad (p^*)^2=M^{*2},\qquad
q^*+p^*=P,
\end{align}
where the last equality uses the reality of $P$. Although complex conjugation
appears in the transformation of the momentum labels, $R$ acts linearly, not
antilinearly, on state
superpositions. Since
$R$ permutes degenerate pair states, $[R,\tilde{\mathbf H}]=0$ on the pair subspace.

In the CoM frame, $R$ is equivalent to interchanging
$\mathbf k\leftrightarrow-\mathbf k$. It is also equivalent to interchanging
the masses $M\leftrightarrow M^{*}$ while holding everything else fixed.

To calculate inner products, we implicitly use the symmetrized tensor-product
representation
\begin{align}
    |M_p M_q^*\rangle
    = \frac{1}{\sqrt{2}}(|M_p\rangle \otimes |M_q^*\rangle
    +|M_{q}^*\rangle \otimes |M_p\rangle).
\end{align}
The swap symmetry separates the pair space into its even and odd subspaces. The
following superpositions are $R$-eigenstates with eigenvalues $+1$ and $-1$,
\begin{align}
|MM^*\rangle^\pm_{\mathbf P,\mathbf k}
=\frac{1}{\sqrt{2}}(|M_p M^*_q\rangle
\pm |M_{q^*} M^*_{p^*}\rangle).
\label{epm}\end{align}
The only nonzero contributions to the norm come from the inner products between the two
components of the superposition,
\begin{align}
\langle M_p M^*_q|M_{q^*} M^*_{p^*}\rangle
&=
\langle M_p|M^*_{p^{*}}\rangle
\langle M^*_{q}|M_{q^{*}}\rangle,\\
\langle M_{q^*} M^*_{p^*}|M_p M^*_q\rangle
&=
\langle M^*_{p^{*}}|M_{p}\rangle
\langle M_{q^{*}}|M^*_{q}\rangle,
\end{align}
which yield
\begin{align}
{}^\pm_{\mathbf P,\mathbf k}\langle MM^*| MM^*\rangle^\pm_{\mathbf P,\mathbf k}
&= \pm|\langle M^*|M\rangle|^2,\\
 {}^\pm_{\mathbf P,\mathbf k}\langle MM^*| MM^*\rangle^\mp_{\mathbf P,\mathbf k}&=0.
\end{align}

Because the pair state $|MM^*\rangle^\pm_{\mathbf P,\mathbf k}$ is entangled,
it can propagate according to the $MM^{*}$ pair propagator identified in the
previous section, namely the product of the propagators in (\ref{e5}). This
propagator causes a transition between the two components of the superposition.

We obtain the one-pair completeness relation by first simplifying the notation,
\begin{align}
|1\rangle=|M_p M^*_q\rangle,\qquad
|2\rangle=|M_{q^*} M^*_{p^*}\rangle,\qquad
\langle1|2\rangle=\langle2|1\rangle
=|\langle M^*|M\rangle|^2.
\label{e4}\end{align}
Since $\langle1|1\rangle=\langle2|2\rangle=0$, the identity and the swap operator on
this two-state subspace are
\begin{align}
\bm1_{\rm pair}
=\frac{|1\rangle\langle2|+|2\rangle\langle1|}{\langle1|2\rangle},\qquad
R_{\rm pair}
=\frac{|1\rangle\langle1|+|2\rangle\langle2|}{\langle1|2\rangle}.
\end{align}
The projector onto the positive-norm, $R$-even state is consequently
\begin{align}
\frac{1}{2}\left(\bm1_{\rm pair}+R_{\rm pair}\right)
=\frac{1}{2}\frac{(|1\rangle+|2\rangle)(\langle1|+\langle2|)}{\langle1|2\rangle}
=\frac{|MM^*\rangle^+_{\mathbf P,\mathbf k}
{}^+_{\mathbf P,\mathbf k}\langle MM^*|}{\langle1|2\rangle}.
\label{epairprojector}
\end{align}
The corresponding projector onto the negative-norm, $R$-odd state is
\begin{align}
\frac{1}{2}\left(\bm1_{\rm pair}-R_{\rm pair}\right)
=-\frac{|MM^*\rangle^-_{\mathbf P,\mathbf k}
{}^-_{\mathbf P,\mathbf k}\langle MM^*|}{\langle1|2\rangle},
\end{align}
where the minus sign accounts for its negative norm. With
$|MM^*\rangle^\pm_{\mathbf P,-\mathbf k}
=\pm|MM^*\rangle^\pm_{\mathbf P,\mathbf k}$, the projectors are unchanged
under $\mathbf k\leftrightarrow-\mathbf k$, and thus integration over
all directions of $\mathbf k$ requires a symmetry factor of
$1/2$. The two projectors onto the $R=\pm1$ states for fixed $\mathbf P$ are
\begin{align}
\bm1^{(1\,\mathrm{pair})}_{{\rm phys},R=\pm1}(\mathbf P)
&=\pm\frac{1}{2}\int d^3\mathbf k\,
{\cal N}\,
\frac{|MM^*\rangle^\pm_{\mathbf P,\mathbf k}
{}^\pm_{\mathbf P,\mathbf k}\langle MM^*|}{\langle1|2\rangle}.
\label{ephysicalcompleteness}
\end{align}
Their sum gives the one-pair completeness relation.
Here ${\cal N}(\mathbf P,\mathbf k)$ is the normalization factor associated with
using $\mathbf k$ as the continuous variable; it is even under
$\mathbf k\leftrightarrow-\mathbf k$. We determine this factor
in Section~\ref{s6}.

A multipair state is a direct product of single pairs, and its norm is the
product of the single-pair norms.
The requirement that each pair has a real total 4-momentum fixes its
constituent pairing for generic multipair kinematics. In particular, $R$
acts only within each such pair and does not interchange constituents
between different pairs, since such an interchange would generally make
the corresponding pair momenta complex. Thus, $R$ acts as the product of the
single-pair swap operations, and the $R$ eigenvalue of the state is the product
of the eigenvalues of its constituent pairs.

The main point of this section is to impose a second superselection rule on the
theory, this time defined by $R$.
Both the $R=+1$ and $R=-1$ sectors remain physical. States in the $R=+1$
($R=-1$) sector contain an even (odd) number of $R=-1$ pairs, in addition to
any number of $R=+1$ pairs. A state with no pairs is in the $R=+1$ sector.
Transitions between the two sectors and superpositions across them are
forbidden. These restrictions ensure positive probabilities through the same
mechanism described for the $Q$ superselection rule in \cite{realsp}.

\section{Pair-based old-fashioned perturbation theory}\label{s5}

Old-fashioned perturbation theory
(OFPT) can be formulated directly in this pair-based state space. The free
basis consists of normal-particle states and pair states $|MM^*\rangle^\pm_{\mathbf P,\mathbf k}$. Each pair has the real free energy
$\Omega(\mathbf P,\mathbf k)$ defined in (\ref{epairenergy}).
The completeness relation for intermediate states requires integrations over the real
total momentum $\mathbf P$ and the
real internal momentum $\mathbf k$. If an interaction vertex transfers
a real spatial momentum $\mathbf r$ to a pair, its new total momentum
$\mathbf P'=\mathbf P+\mathbf r$ remains real. The resulting intermediate
state is expanded over allowed real internal momenta $\mathbf k'$ and has the
real energy $\Omega(\mathbf P',\mathbf k')$.

For any time ordering, an intermediate state containing normal particles
indexed by $a$ and pairs indexed by $j$ has the energy denominator
\begin{align}
\frac{i}{E_{\rm in}-\sum_a E_a-
\sum_j\Omega(\mathbf P_j,\mathbf k_j)+i\epsilon}.
\label{eofptgeneral}
\end{align}
All energies in this expression are real. Thus, if pair-based OFPT is taken as
the direct definition of perturbation theory for the superselected theory, no
loop-energy contours around complex constituent poles are required.

The usual $i\epsilon$ in (\ref{eofptgeneral}) is appropriate at
intermediate-state thresholds that respect both $Q$ and $R$
superselection. The free and interacting parts of the full Hamiltonian typically do not
respect $Q$, and so a transformed perturbation theory is needed, as discussed for $Q$
superselection in \cite{realsp,pertur}. This can generate intermediate-state
thresholds that do not use the $i\epsilon$ prescription. The present
construction also requires the projection $Q^{2}|s\rangle=|s\rangle$. With
this projection, the resulting perturbation theory may respect $R$ without
any further transformation.

\subsection{Optical theorem and cutting rules}

Consider an $R$-even pair state $|MM^*\rangle^+_{\mathbf P,\mathbf k}$ in the
intermediate state in (\ref{eofptgeneral}). Let
$E'_{\rm in}=E_{\rm in}-\sum_a E_a$ denote the initial energy minus the energies of
normal particles in the intermediate state. Spatial
momentum conservation fixes the real total pair momentum $\mathbf P$. The
corresponding denominator for a single pair is
\begin{align}
\frac{i}{E'_{\rm in}-\Omega(\mathbf P,\mathbf k)+i\epsilon}.
\end{align}

After using the spatial momentum delta function at each vertex, an integral
over the internal pair momentum $\mathbf k$ remains. The corresponding
contribution has the form
\begin{align}
{\cal I}_{\rm pair}(E'_{\rm in},\mathbf P)
=\int d^3\mathbf k\,
{\cal N}
\frac{i}{E'_{\rm in}-\Omega(\mathbf P,\mathbf k)+i\epsilon}\,
{\cal F}(\mathbf P,\mathbf k).
\label{eofptpair}
\end{align}
In defining ${\cal I}_{\rm pair}$, the external momenta and all other independent
intermediate-state momenta are held fixed. The quantities $E'_{\rm in}$ and
$\mathbf P$ may depend on those momenta, and any remaining integrations over
them are implicit. The normalization ${\cal N}(\mathbf P,\mathbf k)$ was
introduced in (\ref{ephysicalcompleteness}) and is specified in the next
section. The function ${\cal F}(\mathbf P,\mathbf k)$ denotes the remaining
factors in the amplitude. The discontinuity of (\ref{eofptpair}) is
\begin{align}
\mathrm{Disc}\,{\cal I}_{\rm pair}(E'_{\rm in},\mathbf P)
=2\pi\int d^3\mathbf k\,
{\cal N}
\delta(E'_{\rm in}-\Omega(\mathbf P,\mathbf k))\,
{\cal F}(\mathbf P,\mathbf k).
\label{ediscpair}
\end{align}
On the support of this delta function, ${E'_{\rm in}}^{2}-\mathbf P^2=\mu^2(|\mathbf k|)$.
We then define the pair phase-space measure at fixed total spatial momentum as
\begin{align}
 d\Pi_{MM^*}(E'_{\rm in},\mathbf P)
= 2\pi d^3\mathbf k\,
{\cal N}\,
\delta(E'_{\rm in}-\Omega(\mathbf P,\mathbf k)).
\label{epairphase}
\end{align}
We simplify this in the next section.

While the support of the cut is determined by (\ref{epairphase}),
${\cal F}(\mathbf P,\mathbf k)$ contains the residue factor associated with the
$MM^{*}$ pair propagator. In OFPT, this factor is
\begin{align}
{\cal R}_{MM^*}(\mathbf k)
=\left|
\frac{\langle  0|\phi|M\rangle
\langle M^*|\phi| 0\rangle}
{\langle M^*|M\rangle}
\right|^2.
\label{epairres}
\end{align}
Denoting the production amplitude for
$|MM^*\rangle^+_{\mathbf P,\mathbf k}$ by $A(\mathbf k,\mathbf P)$, the pair
contribution to the cut side of the optical theorem is therefore
\begin{align}
\int d\Pi_{MM^*}(E'_{\rm in},\mathbf P)\,
|A(\mathbf k,\mathbf P)|^2\,{\cal R}_{MM^*}(\mathbf k).
\label{ecutpair}
\end{align}
Dependence on and integrations over any other final-state momenta are implicit.

On the other side of the optical theorem, the on-shell intermediate-state sum
uses the pair phase-space measure identified from the discontinuity of
(\ref{eofptpair}). Inserting the completeness relation from
(\ref{epairprojector}) and the extra $1/2$ in (\ref{ephysicalcompleteness}) gives
\begin{align}
\frac{1}{4}\int d\Pi_{MM^*}(E'_{\rm in},\mathbf P)\,
\frac{1}{\langle1|2\rangle}\bigl(&
\langle i|T^\dagger|1\rangle\langle1|T|i\rangle
+\langle i|T^\dagger|1\rangle\langle2|T|i\rangle\nonumber\\
&+\langle i|T^\dagger|2\rangle\langle1|T|i\rangle
+\langle i|T^\dagger|2\rangle\langle2|T|i\rangle\bigr).
\label{eotpairprojected}
\end{align}
For an $R$-even initial state and an
$R$-invariant transition operator,
$\langle 1|T|i\rangle=\langle2|T|i\rangle$ and
$\langle i|T^\dagger|1\rangle=\langle i|T^\dagger|2\rangle$.
The sum of four terms in (\ref{eotpairprojected}) reduces to
\begin{align}
\int d\Pi_{MM^*}(E'_{\rm in},\mathbf P)\,
\frac{\langle i|T^\dagger|1\rangle
\langle2|T|i\rangle}{\langle1|2\rangle}.
\label{eotpairnew}
\end{align}
LSZ reduction gives
\begin{align}
\langle 2|T|i\rangle
&=A(\mathbf k,\mathbf P)
\langle M|\phi|0\rangle
\langle M^*|\phi|0\rangle,\\
\langle i|T^\dagger|1\rangle
&=A(\mathbf k,\mathbf P)^*
\langle0|\phi|M\rangle
\langle0|\phi|M^*\rangle.
\end{align}
With $\langle1|2\rangle$ given in (\ref{e4}), these expressions reproduce
(\ref{ecutpair}), thereby recovering the optical theorem. The entanglement of
the pair state is needed on the left-hand side of the optical theorem to
produce pair propagation and on the right-hand side to produce a positive-norm
state from the superposition of $|1\rangle$ and $|2\rangle$.

\section{Normalization, phase space, and self-energy}\label{s6}

\subsection{Normalization factor}

We fix ${\cal N}$ by matching to pair propagation in the CoM frame, where the
constituent spatial momenta are the real vectors $\mathbf k$ and $-\mathbf k$.
After factoring out the pole residues, the two positive-energy residues in the
product of the propagators in (\ref{e5}) give the CoM kinematic measure
\begin{align}
{\cal N}(\mathbf0,\mathbf k)
=\frac{1}{(2\pi)^3}\frac{1}{4E_{\mathbf k}E_{\mathbf k}^*}.
\end{align}
For the real total pair momentum $\mathbf P$, covariant normalization gives the measure
$d^3\mathbf P/(2\Omega)$, rather than its CoM value
$d^3\mathbf P/(2\mu)$. The boost from the CoM frame therefore supplies the
factor $\mu/\Omega$, and
\begin{align}
{\cal N}(\mathbf P,\mathbf k)
=\frac{1}{(2\pi)^3}\frac{\mu(|\mathbf k|)}
{4\Omega(\mathbf P,\mathbf k)|E_{\mathbf k}|^2}.
\label{epairnormalization}
\end{align}
In a boosted frame, the resulting constituent spatial momenta are generally
complex, but no delta function of those complex momenta is introduced.

\subsection{Pair phase space}

We can write the pair phase-space measure in (\ref{epairphase}) more explicitly.
With $\mu(|\mathbf k|)=E_{\mathbf k}+E_{\mathbf k}^*$, the derivative needed for the
radial delta function is
\begin{align}
\frac{d\Omega}{dk}=\frac{\mu}{\Omega}\frac{d\mu}{dk},\qquad\frac{d\mu}{dk}=\frac{k\mu}{|E_{\mathbf k}|^2}.
\end{align}
Define $s={E'_{\rm in}}^{2}-\mathbf P^2$ and $k_s$ through
$\mu^2(k_s)=s$. Writing $M=a+ib$, we have
\begin{align}
k_s^2=\frac{\lambda(s,M^2,M^{*2})}{4s}
=\frac{(s-4a^2)(s+4b^2)}{4s},
\label{epairks}
\end{align}
where $\lambda$ is the K\"all\'en function. With the normalization factor ${\cal N}$ in (\ref{epairnormalization}), the pair phase space therefore becomes
\begin{align}
d\Pi_{MM^*}
&=\frac{k_s}{16\pi^2\sqrt{s}}d\Omega_{\mathbf k}
=\frac{\sqrt{(s-4a^2)(s+4b^2)}}{32\pi^2s}
d\Omega_{\mathbf k},\label{epairphasedifferential}
\end{align}
with support for $s\geq4a^2$. Thus, the physical cut places the real total
momentum of the pair on a continuum of mass shells with a positive measure. The
angular integration in (\ref{epairphasedifferential}) remains nontrivial when
the interaction matrix elements depend on the direction of $\mathbf k$.

\subsection{A self-energy bubble}

As a simple application, let $\chi$ be a normal real scalar and consider the
interaction
\begin{align}
{\cal L}_{\rm int}=-g\chi\varphi\varphi^\dagger,
\label{echiinteraction}
\end{align}
where $\varphi$ and $\varphi^\dagger$ were introduced in (\ref{e1}). This
illustrative coupling is neutral under $Q$ and invariant under $R$ and
couples a $\chi$ state to an $R$-even pair.\footnote{A coupling to a single pair is
not produced by the minimal coupling to gravity mentioned in footnote 2. Rather, the
effective coupling there is to a pair of pairs.}

The two time orderings of the pair contribution to the $\chi$ self-energy can
be written as
\begin{align}
\Sigma_{\rm pair}(P^0,\mathbf P)
=g^2|Z|^2\int d^3\mathbf k\,{\cal N}(\mathbf P,\mathbf k)
\left(\frac{1}{\Omega-P^0-i\epsilon}
+\frac{1}{\Omega+P^0-i\epsilon}\right).
\label{echisigmaofpt}
\end{align}
We again include the residue factor for the pair propagator, which we now write as $|Z|^2$. Using (\ref{epairnormalization}), the two denominators combine to give
\begin{align}
\Sigma_{\rm pair}(P^2)
=g^2|Z|^2\int\frac{d^3\mathbf k}{(2\pi)^3}
\frac{\mu(|\mathbf k|)}{2|E_{\mathbf k}|^2}
\frac{1}{\mu^2(|\mathbf k|)-P^2-i\epsilon}.
\label{echisigmainvariant}
\end{align}
Thus, the sum of the time orderings depends only on the Lorentz invariant
$P^2=(P^0)^2-\mathbf P^2$.

Changing variables from $|\mathbf k|$ to
$s=\mu^2(|\mathbf k|)$ gives the spectral representation
\begin{align}
\Sigma_{\rm pair}(P^2)
&=g^2\int_{4a^2}^{\infty}ds\,
\frac{\rho_{MM^*}(s)}{s-P^2-i\epsilon},
\label{echisigmaspectral}\\
\rho_{MM^*}(s)
&=\frac{|Z|^2}{16\pi^2}
\frac{\sqrt{(s-4a^2)(s+4b^2)}}{s}
\theta(s-4a^2),
\label{epairspectraldensity}
\end{align}
where $M=a+ib$. The spectral density is real and positive.
The integral in (\ref{echisigmaspectral}) is
logarithmically divergent, as expected for a scalar bubble in four dimensions.
The finite discontinuity $2\,\mathrm{Im}\,\Sigma_{\rm pair}(s)$
is $g^{2}|Z|^{2}$ times the integral of the pair phase space in
(\ref{epairphasedifferential}).

\section{Conclusion}\label{s7}

This work explores whether complex-conjugate poles on the physical sheet can
participate in a probabilistically consistent $S$-matrix without being removed
from the canonically quantized theory. The construction relies on two
complementary restrictions. The $Q$ superselection rule, together with the
physical-state condition $Q^{2}|s\rangle=|s\rangle$, decouples sectors with
unequal numbers of $M$ and $M^*$ excitations, while the $R$ superselection
rule restricts pairs to entangled $R$-eigenstates and ensures positive
probabilities.
In the optical theorem, the resulting pair state appears as a composite
intermediate excitation; its real total momentum lies on one of a continuum
of pair mass shells, and its completeness relation reproduces the pair propagator
residue. For a cut containing a single $R=-1$ pair state, the cut weight is
negative, but superselection requires the initial state to have $R=-1$ as
well, so the associated probability is positive.

The physical excitation is an entangled $MM^*$ pair whose
constituent momenta and energies are complex but whose total 4-momentum is
real. The two sectors selected by $R$ have states with even and odd numbers of
$R$-odd pairs, respectively.
All these pairs behave as $Q=+1$ composite excitations with a continuously
varying real invariant mass. Thus, although the construction began by relaxing
the real-spectrum constraint, the observable pair spectrum is effectively real.

The projected physical state space is a direct product of two Krein spaces and thus has four
sectors, characterized by even or odd numbers of normal ghosts and even or odd
numbers of $R$-odd pairs.
The existence of a real spectrum in a Krein space allows for an
alternative positive-definite inner product \cite{Mostafazadeh:2001nr}. Such an
inner product was used in \cite{realsp,pertur} to construct a
transformed perturbation theory that makes the $Q$ superselection rule manifest
at each finite order. The result was a modified version of OFPT.
We have found that an OFPT-like approach is also useful here for addressing the
pair sector in superselected ghost theories. In this context, Lorentz
covariance, locality, renormalization, and general multipair cuts deserve
further study.

The advantages of Feynman perturbation theory for conventional
theories do not carry over to ghost theories, so reliance on the Feynman
approach may have hindered progress with these theories.

\section*{Acknowledgements}
Various LLMs have helped to accelerate research of superselected ghost theories.


\begin{thebibliography}{99}

\bibitem{realsp}
B.~Holdom, ``Superselected ghost theory: real spectrum,'' arXiv:2608.06605.

\bibitem{pertur} B.~Holdom, ``Superselected ghost theory: perturbation theory,'' arXiv:2608.09017.

\bibitem{LW:1969}
T.~D.~Lee and G.~C.~Wick, ``Negative metric and the quantization of a field,'' Nucl. Phys. B \textbf{9}, 209--243 (1969); ``Finite theory of quantum electrodynamics,'' Phys. Rev. D \textbf{2}, 1033--1048 (1970).

\bibitem{Grinstein:2007mp}
B.~Grinstein, D.~O'Connell and M.~B.~Wise, ``The Lee-Wick standard model,'' Phys. Rev. D \textbf{77}, 025012 (2008) [arXiv:0704.1845].

\bibitem{Donoghue:2019ecz}
J.~F.~Donoghue and G.~Menezes, ``Unitarity, stability and loops of unstable ghosts,'' Phys. Rev. D \textbf{100}, 105006 (2019) [arXiv:1908.02416].

\bibitem{Anselmi:2018kgz}
D.~Anselmi, ``Fakeons, microcausality and the classical limit of quantum gravity,'' Class. Quant. Grav. \textbf{36}, 065010 (2019) [arXiv:1809.05037].

\bibitem{Kubo:2023lpz}
J.~Kubo and T.~Kugo,
``Unitarity violation in field theories of Lee{\textendash}Wick{\textquoteright}s complex ghost,''
PTEP \textbf{2023}, no.12, 123B02 (2023)
[arXiv:2308.09006 [hep-th]].

\bibitem{Kubo:2024ysu}
J.~Kubo and T.~Kugo, ``Anti-Instability of Complex Ghost,'' PTEP \textbf{2024}, no.~5, 053B01 (2024) [arXiv:2402.15956].

\bibitem{Buoninfante:2026mve}
L.~Buoninfante,
``Asymptotic Quantum Dynamics of Ghost Fields,''
[arXiv:2605.29047 [hep-th]].

\bibitem{Buoninfante:2026jfj}
L.~Buoninfante,
``Ghosts versus Unstable Particles in Quantum Field Theory,''
[arXiv:2606.18349 [hep-th]].

\bibitem{Holdom:2024onr}
B.~Holdom,
``Making sense of ghosts,''
Nucl. Phys. B \textbf{1008}, 116696 (2024)
[arXiv:2408.04089 [hep-th]].

\bibitem{naka}
N.~Nakanishi, ``Covariant Formulation of the Complex-Ghost Relativistic Field Theory and
the Lorentz noninvariance of the S Matrix,'' Phys. Rev. D \textbf{5}, 1968 (1972).

\bibitem{Mostafazadeh:2001nr}
A.~Mostafazadeh, ``Pseudo-Hermiticity versus PT symmetry II: A complete characterization of non-Hermitian Hamiltonians with a real spectrum,'' J. Math. Phys. \textbf{43}, 2814--2816 (2002) [arXiv:math-ph/0110016 [math-ph]].

\end{thebibliography}
\end{document}